\documentclass{INTERSPEECH2023}
\usepackage{multirow}
\usepackage{booktabs}
\usepackage{array}
\usepackage{subcaption}
\interspeechcameraready 
\title{Pretraining Conformer with ASR or ASV for Anti-Spoofing Countermeasure}
\name{Yikang Wang$^{1,2}$, Hiromitsu Nishizaki$^1$, Ming Li$^{2,3}$}%, Hongbin Suo$^4$, Yulong Wan$^4$}
\address{
  $^1$Integrated Graduate School of Medicine, Engineering, and Agricultural Sciences, \\University of Yamanashi, Japan\\
  $^2$Suzhou Municipal Key Laboratory of  Multimodal Intelligent Systems, \\Duke Kunshan University, Kunshan, China\\
  $^3$School of Computer Science, Wuhan University, China}
\email{wwm1995@alps-lab.org, hnishi@yamanashi.ac.jp, ming.li369@duke.edu}

\begin{document}
%\ninept
%
\maketitle
\begin{abstract}
Finding synthetic artifacts of spoofing data will help the anti-spoofing countermeasures (CMs) system discriminate between spoofed and real speech. The Conformer combines the best of convolutional neural network and the Transformer, allowing it to aggregate global and local information. This may benefit the CM system to capture the synthetic artifacts hidden both locally and globally. In this paper, we present the transfer learning based MFA-Conformer structure for CM systems. By pre-training the Conformer encoder with different tasks, the robustness of the CM system is enhanced. The proposed method is evaluated on both Chinese and English spoofing detection databases. In the FAD clean set, proposed method achieves an EER of 0.04\%, which dramatically outperforms the baseline. Our system is also comparable to the pre-training methods base on Wav2Vec 2.0. Moreover, we also provide a detailed analysis of the robustness of different models.
\end{abstract}

\section{Introduction}
Automatic speaker verification (ASV) techniques have become widely used in real-life due to the development of deep neural network \cite{gao2019improving, desplanques20_interspeech, chen2022large}. However, the rapid development of generative techniques, e.g.  text-to-speech synthesis (TTS) and voice conversion (VC), makes the ASV systems vulnerable \cite{mittal2022automatic, shiga2020text}. Hense, a robust anti-spoofing countermeasures (CMs) system, is very important as a safeguard for the ASV systems \cite{das2020attacker,evans2014speaker}.

There are two major challenges in logical access (LA) based anti-spoofing CM tasks \cite{ tan2021survey_asv}. One is the noise robustness problem. In practice, end devices often collect data in complex scenarios that contain a lot of noise, which leads to degraded performance of the CM systems. Another is the problem of unseen spoofing data detection. As the TTS and VC algorithms continue to advance, the CM system has to face unseen data generated by unknown spoofing algorithms, which also leads to low accuracy. 
In this paper, besides using the English-based ASVspoof database \cite{asvspoof19,asvspoof21}, we also focus on the FAD database \cite{ma2022fad}, which consists of Chinese speech data, and separating seen and unseen samples as different subsets in the test set. In this work, we want to build a rubust CM system in noisy and unseen data scenarios.

Recent research works have shown that self-supervised learning (SSL) using large models can learn generalized speech representations from vast amounts of unlabeled data, demonstrating robustness and strong generalization in various speech-related downstream tasks \cite{fan2020exploring}. Models such as Wav2Vec \cite{schneider2019wav2vec}, HuBERT \cite{hsu2021hubert}, Wav2Vec 2.0 \cite{baevski2020wav2vec,xu2021self}, and WavLM \cite{chen2022wavlm} have exhibited promising results in speech recognition \cite{baevski2020wav2vec}, emotion recognition \cite{siriwardhana2020multimodal}, speaker recognition \cite{chen2022large}, and also anti-spoofing CM tasks \cite{wang2021investigating,tak2022automatic}. Wang et al. \cite{wang2021investigating}, compare the performance of CM systems with different combinations of self-supervised pretrained front-ends and various back-ends. Tak et al. \cite{tak2022automatic} discusse the potential to improve generalization and domain robustness through the use of wav2vec 2.0 XLSR as front-end of AASIST \cite{aasist} CM network. They also suggest using telephone channel based data augmentation techniques, such as Rawboost \cite{tak2022rawboost}, to enhance the model robustness. Lee et al. \cite{lee22q_interspeech} investigate the exposure of synthetic artifacts in different feature spaces by taking out the outputs of different layers in a 24-layer Transformer front-end as features.
However, training or fine-tuning models with over 300 million parameters is extremely time-consuming and requires large scale computing resources. The MFA-Conformer model effectively integrates global and local information \cite{gulati2020conformer,zhang2022mfa}, and has great potential to be used in robust ASV or anti-spoofing CM tasks. Since the Conformer model has been widely used for ASR , we can easily perform transfer learning on ASR models pre-trained on large amount of data \cite{cai2022pretrain}. In addition, the Conformer model has a much smaller model size compared to other front-end big models, e.g. WavLM \cite{chen2022wavlm}.
\begin{figure*}[t]
  \centering
  \includegraphics[width=1\linewidth]{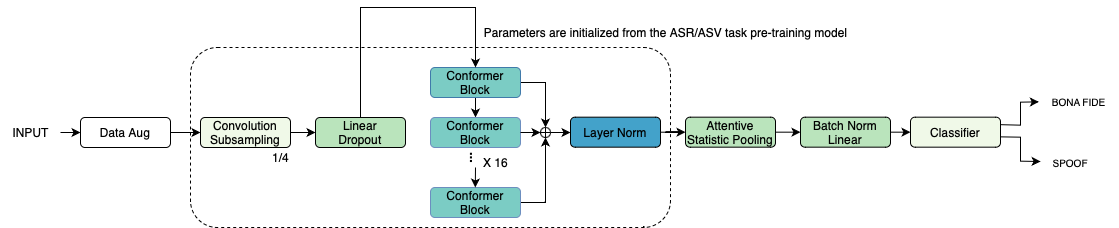}
  \caption{The proposed transfer learning based MFA-Conformer structure for CM systems.}
  \label{fig:approach}
\end{figure*}
In this paper, we propose a transfer learning based MFA-Conformer structure for CM systems. We first pre-train the MFA-conformer model in the ASR or ASV task, and then trim the model to obtain the encoder part, which is fine-tuned on different anti-spoofing databases. Experimental results show that the MFA-Conformer model trained using the proposed transfer learning method achieves better detection performance in both seen and unseen conditions compared to other models in the control group. The main contributions of this paper are summarized as follows:

\begin{itemize}
% \item
% This paper introduces the Conformer model into the anti-spoofing CM task using a transfer learning approach.
\item
We introduce the MFA-Conformer model into the anti-spoofing CM task. Results demonstrate the effectiveness of transfer learning with ASR or ASV pre-trained models.
\item
We analyze the robustness of different CM models against specific spoofing algorithms, propose the error-prone tendency (ET) as a judging metric, and visualize the results in line graphs. This may help the choice of model fusion or help in selecting appropriate features for spoofing algorithm traceability tasks.

\end{itemize}

\section{Method}

\subsection{Conformer architecture}
\label{section:Conformer}

This section describes the main structure of the Conformer encoder used for ASR \cite{gulati2020conformer}, as well as the MFA-Conformer for ASV tasks \cite{zhang2022mfa}. In this paper, the original Conformer and MFA-Conformer structures are pre-trained in the ASR and ASV tasks, respectively, and subsequently fine-tuned for different databases after removing the redundant parts of the model and connecting its encoder to the backend as a trainable feature extractor as shown in Figure~\ref{fig:approach}.
\vspace{-3mm}
\subsubsection{Conformer encoder for ASR}
To learn both position-wise local features and content-based global interactions, Gulati et al. \cite{gulati2020conformer} proposed the Conformer model. In this network, the input audio signals are processed through feature extractor and a subsampling convolution layer and then fed into the Conformer encoder composed of multiple Conformer blocks. A Conformer block is a sandwich structure of four modules stacked together, i.e., a feed-forward (FFN) module, a multi-headed self-attention module (MHSA), a convolution module, and a second FFN module in the end.

The MHSA is employed with a relative sinusoidal positional encoding scheme from Transformer XL \cite{dai2019transformer} to improve generalization on varying input length. The subsequent convolution module contains a point-wise convolution and a gated linear unit (GLU) activation layer, followed by a single 1-D depth-wise convolution layer with batchnorm to training deeper models. Two half-step FFN layers replacing the original FFN layer in the Transformer block, one before the attention layer and one after. In general \cite{gulati2020conformer}, this structure can be represented as
% \[
\begin{align}
    \tilde{x_{i}} & =x_{i}+\frac{1}{2} \operatorname{FFN}\left(x_{i}\right) \\
    % \label{equation:eq3}
    x_{i}^{\prime} & =\tilde{x_{i}}+\operatorname{MHSA}\left(\tilde{x_{i}}\right) \\
    % \label{equation:eq4}
    x_{i}^{\prime \prime} & =x_{i}^{\prime}+\operatorname{Conv}\left(x_{i}^{\prime}\right) \\
    % \label{equation:eq5}
    y_{i} & =\operatorname{Layernorm}\left(x_{i}^{\prime \prime}+\frac{1}{2} \operatorname{FFN}\left(x_{i}^{\prime \prime}\right)\right)
    % \label{equation:eq6}
\end{align}

where FFN refers to the feed forward module, MHSA refers to the Multi-Head Self-Attention module, and Conv refers to the convolution module.
\vspace{-3mm}
\subsubsection{MFA-Conformer encoder for ASV}
\label{section:mfa}
Chen et al. \cite{chen2022large} found that superimposing the output of each Transformer block after hidden layers during pretraining can produce better representations for speaker recognition tasks compared to simply using the output of the last Transformer block layer. Similarly, Zhang et al. \cite{zhang2022mfa} verified this conclusion on the Conformer model and proposed the Multi-scale Feature Aggregation Conformer (MFA-Conformer). This network integrates information from multiple Conformer blocks and connects the outputs of multiple scales, which are then normalized and pooled together to obtain speaker embeddings. In short, the MFA-Conformer is a concatenation of the output of each layer of the original Conformer, leaving the construction of each layer unchanged.

\subsection{Transfer learning strategy}
For the Conformer pretrained models with two different target tasks, we adopt two subtly different transfer learning strategy. Specifically, we load the parameters of the pretrained Conformer encoder into the MFA-Conformer encoder of our CM system, the output of each Conformer block is concatenated to extract embeddings, and an attentive statistics pooling (ASP) layer and a fully-connected (FC) layers are connected to obtain segment-level features, then a linear classifier is connected at the back-end for fine-tuning. The overall system structure is shown in Figure~\ref{fig:approach}. During the transfer learning based fine-tuning process, we do not always freeze the MFA-Conformer as a pure front-end feature extractor; instead, we first fix the parameters of the Conformer encoder and update the parameters for a few training epochs. Then, we jointly fine-tune the Conformer encoder and the linear layer classifier as a whole during training for the anti-spoofing CM tasks.
\vspace{-3mm}
\section{Experimental setup}

\subsection{Data preparation}
\label{section:data}
The contents of the databases used in this paper are summarized in Table~\ref{tab:database}. The fake audio detection (FAD) database \cite{zhang2022mfa} is a Chinese-mandarin database for anti-spoofing CM tasks. It was built to investigate the robustness of spoofing detection methods under noisy conditions. The FAD database has two versions: a clean version and a noisy version. Both versions are divided into different training, development, and test sets in the same way, with no overlap of speakers between the three subsets. Each test set is further divided into seen and unseen subsets. The unseen subset can evaluate the generalization of the CM system to unknown spoofing methods. 
For data augmentation of experiments on FAD database, we used an on-the-fly data augmentation method \cite{cai2020fly}, which is more diverse and efficient. Specifically, when loading audio data, we randomly selected 2/3 of the data in each minibatch for noise addition. The data augmentation method includes adding background noise (environmental noise, music, babble noise) and adding convolutional reverberation. The two augmentation methods use the MUSAN database \cite{musan2015} and the room impulse responses (RIR) database \cite{ko2017study}, respectively. During the on-the-fly data augmentation, the method is randomly selected, and the signal-to-noise ratio is randomly set in the 0 to 20 dB range.

The ASVspoof database is a series of data from the ASVspoof challenges \cite{asvspoof15,asvspoof17,asvspoof19,asvspoof21}. For the experiments, we use the ASVspoof 2019 LA (19LA) \cite{asvspoof19} and the ASVspoof 2021 LA (21LA) \cite{asvspoof21} datasets. The 19LA dataset was created using utterances from 107 speakers (46 male, 61 female). The set of 107 speakers is partitioned into three speaker-disjoint sets for training, development, and evaluation. The spoofed utterances were generated using four TTS and two VC algorithms in the training and development sets, while 13 TTS/VC algorithms are used in the evaluation set, 4 of which are partial-known and 7 of which are unknown for training and development. The 21LA dataset remains the training and development data unchanged and only proposes a new evaluation set that contains attacks using the same simulation methods as the 19LA evaluation set. The 21LA evaluation set consists of the data in various telephone transmission systems, including Voice over Internet Protocol (VoIP) and the Public Switched Telephone Networks (PSTN), thus exhibiting real-world signal transmission channel effects. The experiments using ASVspoof database in this paper are conducted using the 19LA train as the training set, and the three models with the lowest loss in the 19LA development set are selected to be tested in the respective evaluation sets of 19LA and 21LA dataset. 
Since the 19LA dataset consists of clean data, for the fairness of the comparison, none of our experiments in the ASVspoof database use any data augmentation in fine-turning CM models.

\begin{table}[t]
\caption{The data distribution of each database.}
\label{tab:database}
\resizebox{\columnwidth}{!}{%
\begin{tabular}{lccccc}
\toprule
\multirow{2}{*}{} & \multicolumn{3}{c}{\textbf{ASVspoof 2019 LA}} & \multicolumn{2}{c}{\textbf{ASVspoof 2021 LA}} \\ \cmidrule(lr){2-4}\cmidrule(lr){5-6} 
          & train  & dev    & evaluation &\multicolumn{2}{c}{evaluation}   \\ \midrule
bona fide & 2,580  & 2,580  & 7,355 &\multicolumn{2}{c}{14,816}      \\
spoof     & 22,800 & 22,296 & 63,882 &\multicolumn{2}{c}{133,360}     \\ \midrule
          & \multicolumn{5}{c}{\textbf{FAD clean/noisy data}} \\ \cmidrule(lr){2-6}
          & \multicolumn{2}{c}{train}  & dev     & test~seen & test~unseen \\ \midrule
bona fide      & \multicolumn{2}{c}{12,800} & 4,800 & 14,000     & 7,000        \\
spoof      & \multicolumn{2}{c}{25,600} & 9,600  & 28,000     & 14,000       \\ \bottomrule
\end{tabular}%
}\vspace{-3mm}
\end{table}

\subsection{Model pretraining}
\subsubsection{Pretrained ASR Conformer model}
 We use the NEMO STT En Conformer-CTC Small model version 1.0.0 as the pretrained ASR model \footnote{https://catalog.ngc.nvidia.com/orgs/nvidia/teams/nemo/models/\\stt\_en\_conformer\_ctc\_small}, which has the same model structure as \cite{gulati2020conformer} but replaces the Conformer transducer in \cite{gulati2020conformer} with a linear decoder backend and links a connectionist temporal classification (CTC) for decoding. According to the open source code \footnote{https://github.com/NVIDIA/NeMo/blob/2ef544ffe6daa80d38d\\0f494a7e42adcac50a4b9/examples/asr/conf/ssl/conformer/conformer\_ssl.yaml}, the convolutional layer of the Conformer-CTC small model has a downsampling rate of 1/4. The encoder part has 13M parameters with 4 attention heads and 16 conformer blocks. The feature dimension of the encoder convolutional layer is 176, and the feature dimension of the FFN is 704. The model was trained on a composite database called NeMo ASRSET 1.4.1, which comprises more than 34,000 hours of English speech \cite{cai2022pretrain,kuchaiev2019nemo}.

\subsubsection{ASV Conformer pretraining }
For the pre-training of MFA-Conformer with ASV as the target task, we adopt the identical Conformer encoder construction as the NEMO encoder. However, we concatenate the output of each Conformer blocks to create the output embedding. The development set of VoxCeleb 2 \cite{nagrani2020voxceleb} is utilized to train the model from scratch. The training data is collected from 5,994 speakers and contains 1,092,009 speech files. To augment the training data, we employ velocity perturbation techniques, which involve modifying the original database. Specifically, we accelerate the speech by a factor of 1.1 and decelerate them by a factor of 0.9, resulting in two additional versions of each recording. Consequently, the training dataset expand to include 17,982 speakers and a total of 3,276,027 utterances. During the ASV model training, we refer to the hyperparameter settings of Cai et al. \cite{cai2022pretrain} and finally obtain speaker verification results on VoxCeleb 1 consistent with its description in the article.
\vspace{-2mm}
\subsection{Network setup and evaluation metrics}
 
We used log Mel-Filter Bank energy (FBANK) as the acoustic feature in Conformer based experiments. We extract Fast Fourier Transform (FFT) spectrograms with a window length of 1024 and a hop length of 128, using Blackman windows. The number of Mel-filters in FBANK is set to 80 dimensions. 

During fine-tuning on the FAD database, speech samples are truncated or repeated up to 8 seconds before being loaded into the network for CM task fine-tuning. We used Cross-Entropy softmax as the loss function. AdamW is used as the optimizer with an initial learning rate of 0.001. We apply a cosine annealing learning rate scheduler and a 4000-step warm-up strategy. Each experiment is performed using one NVIDIA RTX A6000 GPU, and we set the batch size to 256, with 100 epochs of training per model.

While fine-tuning on the ASVspoof databases, speech samples are truncated or repeated up to 5 seconds. We then reduce the batch size to 64, set the initial learning rate to 0.0001, and increase the dropout by 50\% in the FC layer after the ASP to prevent model overfitting. For the control group model, we use the code and hyperparameter settings mentioned in \cite{ma2022fad,he2016deep,tak2022automatic}. The final evaluation results are averaged over the selected epochs for each model, which are obtained from the epochs in the development subset with the top-3 lowest loss.

\begin{table*}[t]
\caption{The EERs (\%) of each CM system trained with different database on different evaluation set. The best performance among Conformer models is shown in \textit{italics}, and the best performance among all models is shown in \textbf{bold}. The performance is reported as ``average(best)'' from the Top-3 models. $\dagger$ Note that RawBoost was not utilized in all three databases when reproducing the W2V-AASIST models, as telephone transmission coding was involved in the FAD database and 19LA dataset.}
\label{tab:wovad}
\resizebox{\textwidth}{!}{%
\begin{tabular}{llllllllll}
\toprule
\multicolumn{1}{l}{\multirow{3}{*}{\textbf{Model}}} &
  \multirow{3}{*}{\textbf{pre-trained}} &
  \multicolumn{4}{c}{\textbf{FAD database}} &
  \multicolumn{4}{c}{\textbf{ASVspoof database}} \\ \cmidrule(lr){3-6}\cmidrule(lr){7-10}
\multicolumn{1}{c}{} &              & \multicolumn{2}{c}{clean test} & \multicolumn{2}{c}{noisy test} & \multicolumn{2}{c}{19LA} & \multicolumn{2}{c}{21LA}    \\ \cmidrule(lr){3-4}\cmidrule(lr){5-6}\cmidrule(lr){7-8}\cmidrule(lr){9-10}

\multicolumn{1}{c}{} &            & SEEN          & unSEEN         & SEEN           & unSEEN        & SEEN          & unSEEN         & SEEN           & unSEEN  \\ \cmidrule(lr){1-6}\cmidrule(lr){7-10}

LFCC-GMM\cite{ma2022fad}     & -   & 6.47          & 31.90          & 29.79          & 30.31         & -          & -         & -           & -      \\

ResNet34             & $\times$   & 0.13(0.13) & 25.98(25.91)  & 16.7(16.68) & 37.19(37.09) & 2.36(2.27) & 1.86(1.77)  & 16.31(16.06)  & 12.97(12.93) \\ \cmidrule(lr){1-6}\cmidrule(lr){7-10}

W2V-AASIST\cite{tak2022automatic}$\dagger$    & $\surd$   & 0.08(0.06) & 25.63(25.32) & \textbf{2.15(2.03)} & \textbf{26.79(26.08)} & \textbf{0.21(0.17)} & \textbf{0.60(0.42)} & 3.32(1.75) & 7.97(5.41) \\ \cmidrule(lr){1-6}\cmidrule(lr){7-10}

\multirow{3}{*}{\textbf{Ours}} & $\times$ & 0.30(0.21) & 28.63(26.11) & 4.87(4.11) & 26.88(26.50)& 7.14(6.66) & 7.22(6.81) & 24.69(24.45) & 13.51(12.69)  \\
                      & ASR & \textit{\textbf{0.05(0.04)}} & 27.32(25.74) & 3.46(3.11) & 27.69(26.39) & \textit{0.46(0.44)} & \textit{0.96(0.94)} & \textit{\textbf{2.2(1.74)}} & \textit{\textbf{4.29(4.03)}} \\
                     &  ASV & 0.14(0.10) & \textit{\textbf{25.62(25.01)}} & \textit{3.27(3.04)} & \textit{27.25(26.44)} & 1.04(0.93) & 1.6(1.53) & 5.57(5.31) & 6.53(5.99)   \\  \bottomrule
\end{tabular}%
}
\end{table*}
The system performance is reported using Equal Error Rate (EER). The test data of a certain spoofing algorithm consists of the spoofing speech of that type and all the genuine speech from the evaluation set. To assess the robustness of different models against specific spoofing algorithms, we propose the ET as a metric. The ET metric quantifies the extent of misjudgment for each specific spoofing algorithm by a given model. Its value is derived by regularizing the EER of the model across all spoofing algorithms. Mathematically, it can be expressed as follows:
\begin{align}
ET = \frac{{M_i - \min(M_i)}}{{\max(M_i) - \min(M_i)}}
\label{equation:eq7}
\end{align}
where $M_i$ denotes the EER of model $M$ for a specific algorithm $i$. Here, $i$ belongs to $A07$ to $A19$, which encompasses all spoofing algorithms exist in the ASVspoof evaluation set. An ET value of 1 indicates that the algorithm is most likely to be misclassified by model $M$, while a value of 0 indicates the lowest likelihood of misclassification. It's important to note that the ET metric does not indicate probability.

When evaluating the subset of the ASVspoof database, it is divided into seen and unseen portions based on the spoofing algorithm described in \cite{asvspoof19}. The test data for the seen category includes A16, A19 spoofing speech, and all genuine speech. Conversely, the unseen category consists of A10, A11, A12, A13, A14, A15, A18, and all genuine speech. A07, A08, A09, and A17 are considered as partial-seen, as part of their algorithms were used during the generation of the training data.
\section{Experimental results and analysis}

\subsection{Comparison of CM systems trained with different database}
In addition to the official LFCC-GMM baseline, we have also reproduced two networks as control group models on all databases: a non-pretrained commonly used CM models, FBANK-ResNet34 \cite{nagrani2020voxceleb}; and an SSL pre-trained CM model, W2V-AASIST \cite{tak2022automatic} \footnote{https://github.com/TakHemlata/SSL\_Anti-spoofing}. By comparing the results in Table~\ref{tab:wovad}, it can be observed that the pre-trained model yields a performance improvement over the baseline in all databases. Furthermore, the proposed Conformer models, pretrained by ASR, significantly enhance the performance by a large margin on the FAD clean set and the 21LA dataset.
Moreover, all four CM systems utilizing larger models (including the non-pretrained Conformer model) demonstrate improvements in robustness to noise when compared to the LFCC-GMM and ResNet34 models. Among them, the W2V-AASIST model has the best robustness, which indicates that large size model trained on vast amount of data plays an important role in improving noise robustness. The fact that the Conformer pre-trained model using only 1/12 of W2V-AASIST's parameters, achieves comparable results, illustrates the effectiveness of the proposed transfer learning-based Conformer method. We believe that the small-scale Conformer model can mitigate overfitting and has great potential in the CM task when model distillation is used.
\begin{table*}[t]
\caption{The breakdonw EERs (\%) of the CM system on 19LA evaluation set with different spoofing algorithm.}
\label{tab:breakdown_eer}
\resizebox{\textwidth}{!}{%
\begin{tabular}{llllllllllllllll}
\toprule
\multicolumn{1}{c}{\multirow{2}{*}{System}} & \multicolumn{1}{c}{\multirow{2}{*}{Feature}} & \multicolumn{2}{c}{SEEN}             & \multicolumn{4}{c}{partial SEEN} & \multicolumn{7}{c}{unSEEN}                       & \multicolumn{1}{c}{\multirow{2}{*}{Pooled}} \\ \cmidrule(lr){3-4} \cmidrule(lr){5-8}\cmidrule(lr){9-15}
\multicolumn{1}{c}{}                        & \multicolumn{1}{c}{}                         & A16  & A19                           & A07     & A08    & A09   & A17   & A10   & A11  & A12  & A13  & A14  & A15  & A18   & \multicolumn{1}{c}{}                     \\ \cmidrule(lr){1-2}\cmidrule(lr){3-4}\cmidrule(lr){5-8}\cmidrule(lr){9-15}\cmidrule(lr){16-16}
LFCC-GMM \cite{ma2022fad}                                    & LFCC                                         & 6.31 & 13.94                         & 12.86   & 0.37   & 0     & 7.71  & 18.97 & 0.12 & 4.92 & 9.57 & 1.22 & 2.22 & 3.58  & 8.09                                     \\ \cmidrule(lr){1-2}\cmidrule(lr){3-4}\cmidrule(lr){5-8}\cmidrule(lr){9-15}\cmidrule(lr){16-16}
Res2Net-C \cite{kim2023phase}                                   & CQT+phase                                    & 0.34 & 0.33                          & 0.22    & 2.67   & 0.02  & 1.75  & 0.51  & 0.33 & 0.18 & 0.06 & 0.22 & 0.22 & 1.77  & 0.94                                     \\ \cmidrule(lr){1-2}\cmidrule(lr){3-4}\cmidrule(lr){5-8}\cmidrule(lr){9-15}\cmidrule(lr){16-16}
RawGAT-ST \cite{tak21_rawgat}   & Waveform                                     & 0.67 & 0.62                          & 1.19    & 0.33   & 0.03  & 1.44  & 1.54  & 0.41 & 1.54 & 0.14 & 0.14 & 1.03 & 3.22  & 1.19                                     \\
AASIST \cite{aasist}                                & Waveform                                     & 0.72 & 0.62                          & 0.80     & 0.44   & 0     & 1.52  & 1.06  & 0.31 & 0.91 & 0.10  & 0.14 & 0.65 & 3.40   & 0.83                                     \\
W2V+LGF  \cite{wang2021investigating}                                   & Waveform                                     & 0.11 & 0.17                          & 0.12    & 0.14   & 0.07  & 0.05  & 3.58  & 3.06 & 0.12 & 0.02 & 0.18 & 0.97 & 0.23  & 1.28                                     \\
W2V+AASIST                             & Waveform                                     & 0.02 & 0.30                           & 0.04    & 0.02   & 0     & 0.20   & 0.68  & 0.20  & 0.30  & 0    & 0.06 & 0.23 & 0.42  & \textbf{0.37}                                     \\ \cmidrule(lr){1-2}\cmidrule(lr){3-4}\cmidrule(lr){5-8}\cmidrule(lr){9-15}\cmidrule(lr){16-16}
ResNet34                                    & FBANK                                        & 0.87 & 2.97                          & 0.22    & 0.94   & 0.02  & 4.70   & 0.55  & 0.12 & 0.49 & 0.33 & 0.79 & 1.44 & 4.16  & 2.13                                     \\
Ours (w/o P.)                               & FBANK                                        & 0.53 & 9.34                          & 0.19    & 1.12   & 0.11  & 8.40   & 0.49  & 0.30  & 0.12 & 0.41 & 0.29 & 0.12 & 18.11 & 6.06                                     \\
Ours (ASR P.)                               & FBANK                                        & 0.16 & 0.59                          & 0.10     & 0.45   & 0.06  & 0.72  & 0.41  & 0.51 & 0.06 & 0    & 0.20  & 0.14 & 2.04  & \textbf{0.72}                                     \\
Ours (ASV P.)                               & FBANK                                        & 0.22 & 1.30 & 0.06    & 0.55   & 0.04  & 1.52  & 0.29  & 0.30  & 0.10  & 0.02 & 0.3  & 0.18 & 3.86  & 1.31                                     \\  \midrule
Ours (ASV P.) + W2V-AASIST   & - & 0.06 & 0.30 & 0.02 & 0.12 & 0.02 & 0.26 & 0.26 & 0.20 & 0.06 & 0    & 0.08 & 0.08 & 0.88 & 0.31  $\downarrow$                    \\   
Ours (ASV P.) + Ours (w/o P.)& - & 0.20 & 1.32 & 0.06 & 0.53 & 0.04 & 1.52 & 0.30 & 0.23 & 0.10 & 0.08 & 0.16 & 0.08 & 4.62 & 1.51 $\uparrow$ \\ \midrule
\multicolumn{2}{c}{Spoofing system} &\multicolumn{4}{l}{Features}&\multicolumn{10}{c}{System description}\\ \midrule
\multicolumn{2}{l}{A08: NN-based TTS} &\multicolumn{4}{l}{MCC, F0}&\multicolumn{10}{l}{Vocoder is a neural-source-filter waveform model.}\\ 
\multicolumn{2}{l}{A10: NN-based TTS} &\multicolumn{4}{l}{ FBANK}&\multicolumn{10}{l}{Synthesized audio with speaker information added by WaveRNN vocoder.}\\ 
\multicolumn{2}{l}{A17: NN-based VC} &\multicolumn{4}{l}{MCC, F0}&\multicolumn{10}{l}{Waveform generation is based on a direct waveform modification method.}\\ 
\multicolumn{2}{l}{A18: Non-parallel VC} &\multicolumn{4}{l}{i-vector, MFCC, F0}&\multicolumn{10}{l}{Learning a subspace in the i-vector
space that best discriminates speakers.}\\ 
\multicolumn{2}{l}{A19: Transfer-function-based VC} &\multicolumn{4}{l}{LPCC/MFCC, LPC}&\multicolumn{10}{l}{Conversion is conducted only on active speech frames.}\\ \bottomrule
\end{tabular}
}
\end{table*}

% \begin{figure}[t]
%   \centering
%   \includegraphics[width=1\linewidth]{ET.png}
%   \caption{The ET curve of CM systems in Table~\ref{tab:breakdown_eer} (top), the case where the input feature is a raw waveform (bottom left), and the case where the input feature is FBANK (bottom right).}
%   \label{fig:ET}
% \end{figure}
\begin{figure}
  \centering
  \begin{subfigure}{\linewidth}
    \centering
    \includegraphics[width=\linewidth]{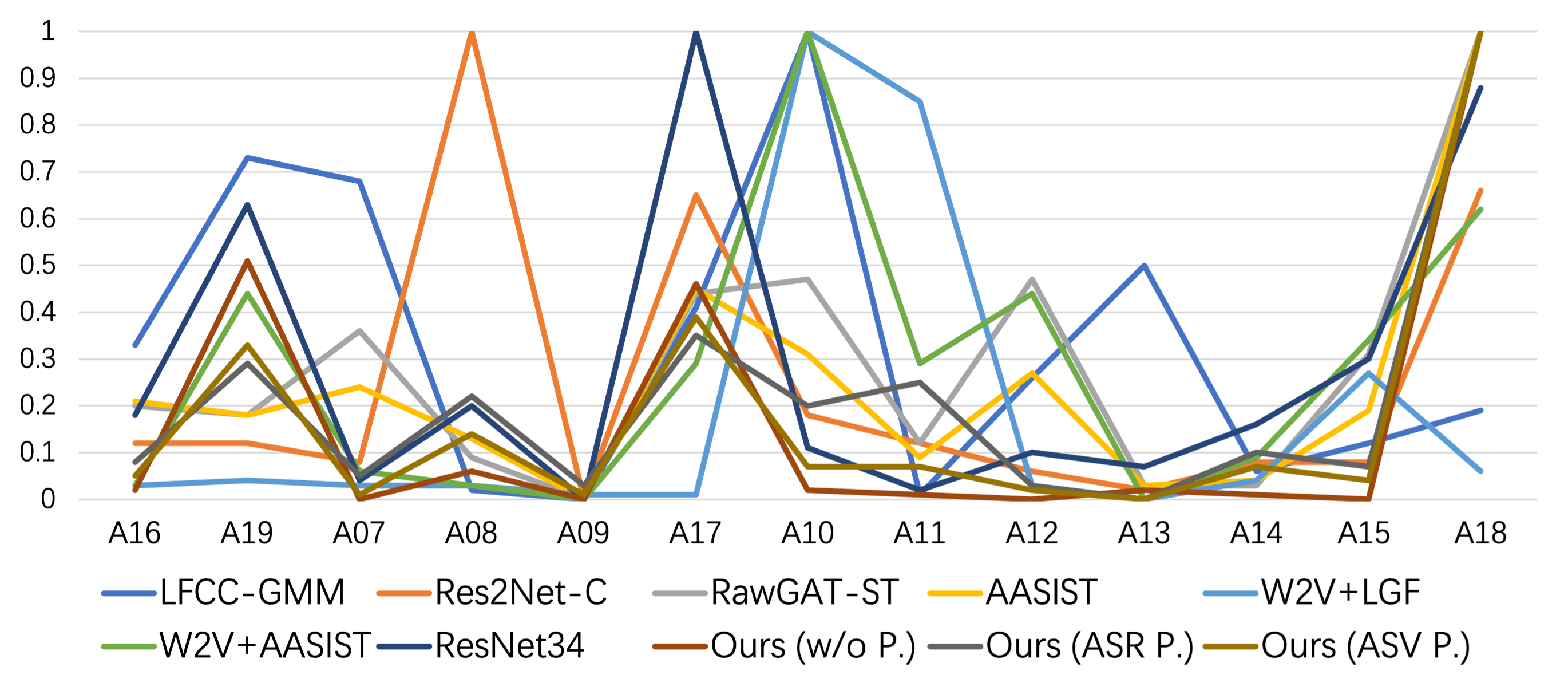}
    \caption{All the CM systems listed in Table~\ref{tab:breakdown_eer}.}
    \label{fig:ET_all}
  \end{subfigure}

  \begin{subfigure}{0.45\linewidth} %0.49
    \centering
    \includegraphics[width=\linewidth]{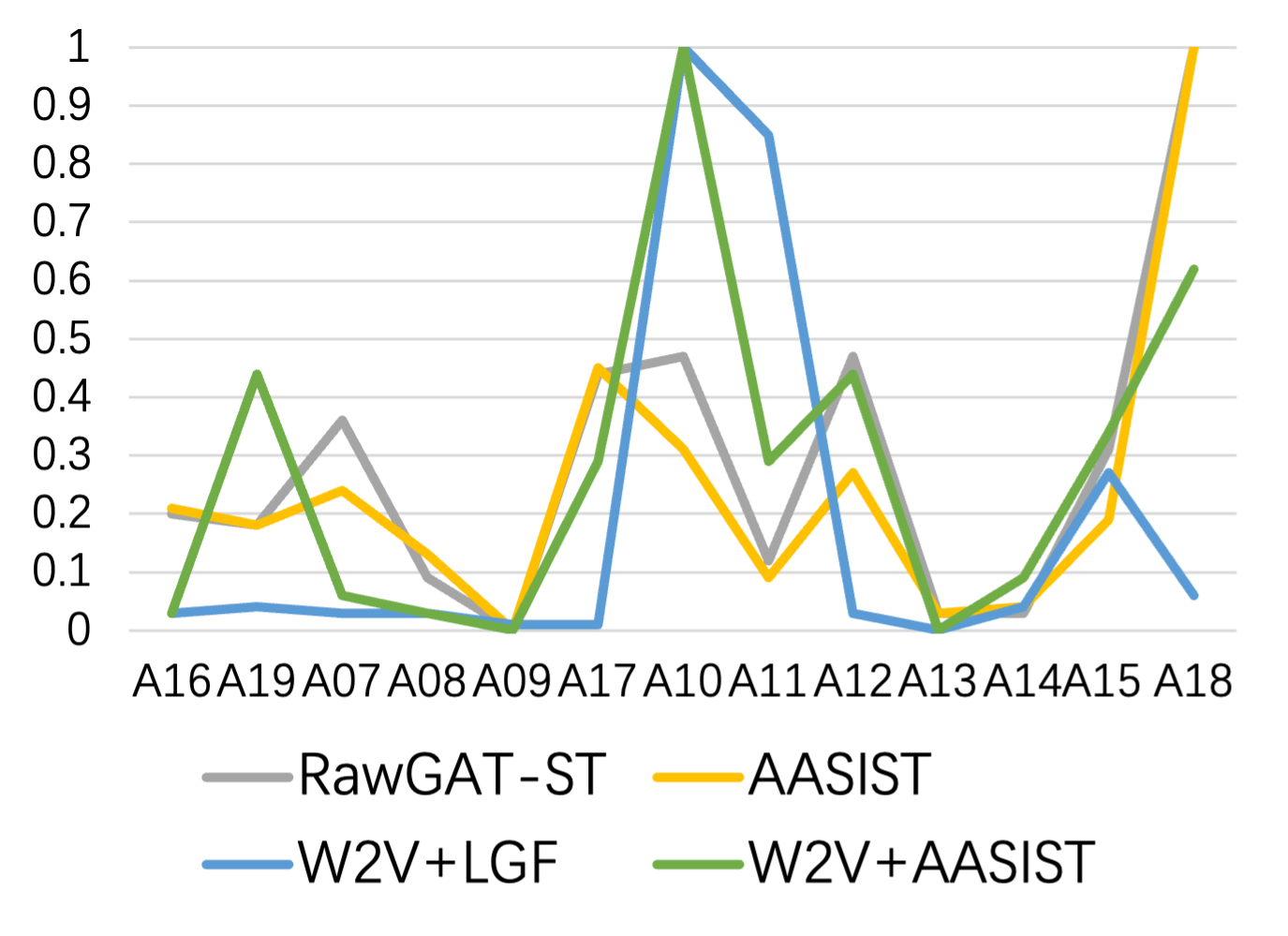}
    \caption{CM system with raw waveform as input.}
    \label{fig:ET_rawwave}
  \end{subfigure}
  \hfill
  \begin{subfigure}{0.45\linewidth} %0.49
    \centering
    \includegraphics[width=\linewidth]{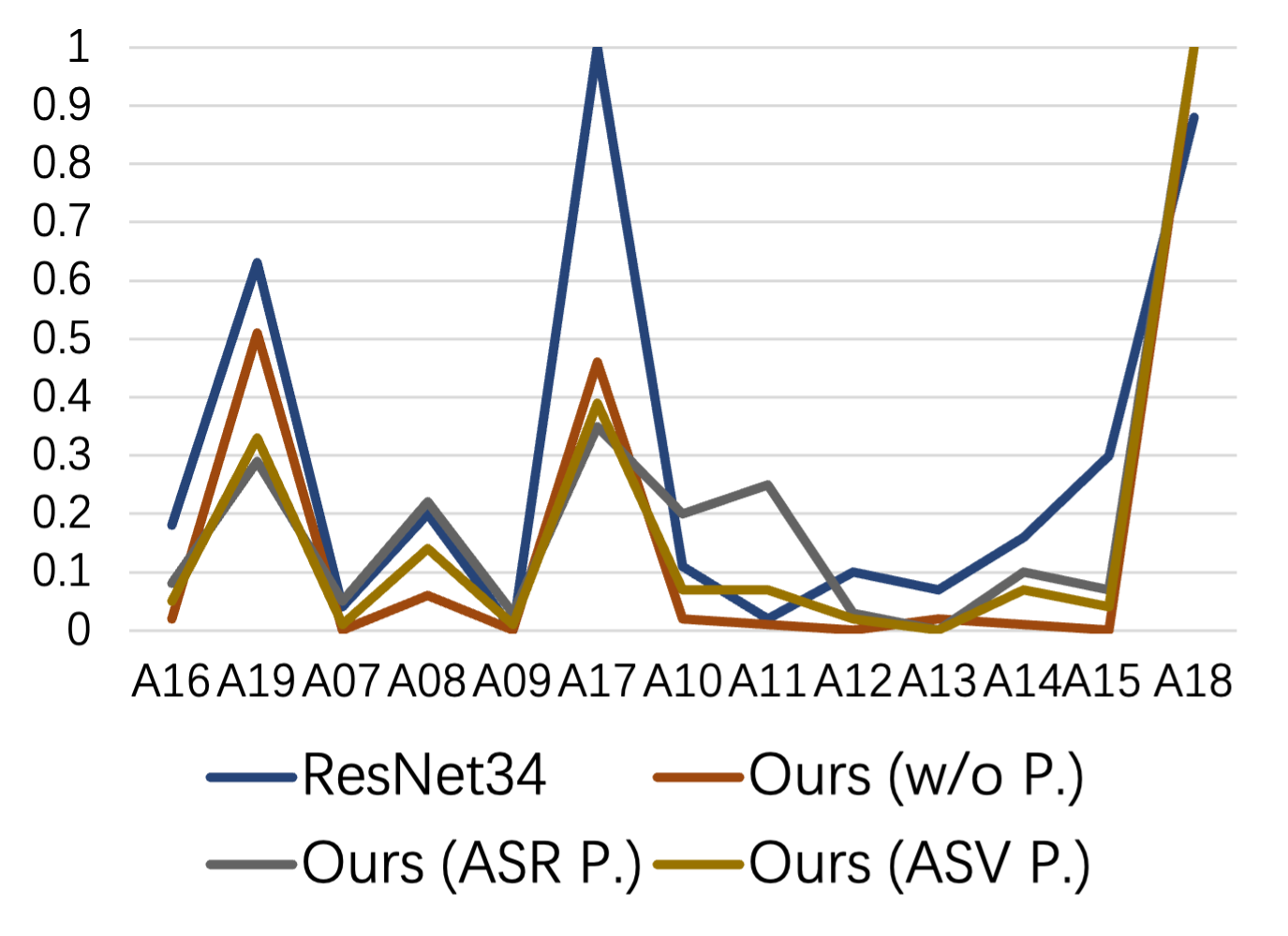}
    \caption{CM system with FBANK as input.}
    \label{fig:ET_FBANK}
  \end{subfigure}

  \caption{The ET curve of CM systems in Table~\ref{tab:breakdown_eer}.\vspace{-5mm}}
  \label{fig:ET}
\end{figure}

\subsection{Analysis of models robustness for certain spoofing algorithm}
We refer to the data generated by the spoofing algorithms that have used in the training set as SEEN, otherwise it is called unSEEN. In Table~\ref{tab:wovad} it is evident that the data obtained from the unSEEN spoofing methods in the FAD database are very difficult to distinguish for any model. However, we can observe a few contrasting conclusions from the ASVspoof database. In particular, the test results of ResNet34 on the 19LA and 21LA datasets, as well as the results of Conformer without pre-training on the 21LA dataset, indicate that the EER of the unSEEN data is lower than that of the SEEN data. To further investigate the reasons behind this, we performed a re-scoring of the model on 19LA dataset, breaking down the scores against spoofing algorithm, and summarized existing breakdown scored system \cite{ma2022fad,kim2023phase,tak21_rawgat,aasist,wang2021investigating} for comparison. From Table~\ref{tab:breakdown_eer}, we can find out that the performance of CM systems is not directly dependent on whether they have encountered a spoofing algorithm before evaluation. For instance, although A19 belongs to SEEN category, the EER obtained by FBANK is only better than that of the LFCC baseline. In the unSEEN category, A11 and A13 are relatively easy to distinguish for each CM system. 

The resulting ET metric for all CM systems were calculated, enabling us to determine which algorithms are more error-prone for different models. The corresponding results are presented in Figure~\ref{fig:ET}, For a specific system, higher values indicate a greater error proneness towards that spoofing algorithm. Since this metric does not represent a probability, it is not feasible to compare the ET values across CM systems. However, it enables us to observe whether different models exhibit the same trend of making mistakes for a certain spoofing algorithm. 
% The \ref{fig} and right curves in Figure~\ref{fig:ET} represent models with raw waveform and FBANK as input features, respectively. 
From Figure~\ref{fig:ET_rawwave}, we observed that the CM systems using raw waveform as input features are easier to misjudge the A10 and A18 spoofing algorithms in unSEEN category. Conversely, from Figure~\ref{fig:ET_FBANK}, systems employing FBANK as input features are more likely to misclassify A19, A08, A17, and A18. It suggest that the FBANK feature-based system exhibits little significant correlation between its error-prone spoofing algorithms and whether or not they were encountered in the training set. 
In Table~\ref{tab:breakdown_eer}, we provide a concise summary of five spoofing algorithms that shown poor performance for most systems, more systems' detail can be found in the paper \cite{asvspoof19}. Based on these descriptions, it can be concluded that the CM system faces difficulties in discriminating data generated by the neural-network-based (NN-based) TTS system when raw waveform features are used as inputs. Furthermore, the CM system is more prone to making mistakes when dealing with VC system-generated data, particularly when FBANK is used as the feature. By referring to Table~\ref{tab:breakdown_eer}, we observe that all systems exhibit significantly high EERs on A18, indicating that the text-independent ASV-based VC system may alter the speaker information to a large extent without leaving noticeable traces of synthesis during the creation of spoofing data. 

When the ET values of two models are similar, the two models have similar tendency to make mistakes, and model fusion may lead to counterproductive results, e.g., ASV-pretrain+w/o pretrain with pooled EER $1.51\uparrow$; whereas, when the ET curves of the two models are more different, the performance enhancement of score fusion is more significant, e.g., ASV-pretrain+W2V-AASIST with pooled EER $0.31\downarrow$.

\subsection{Comparison with state-of-the-art systems}
Table~\ref{tab:19LAeer} presents a comparison of the proposed ASR/ASV pretrained Conformer model to the performance of several recently proposed single model \cite{aasist, tak21_rawgat,zhang21da_interspeech,lee22q_interspeech,tak2022automatic,wang2021investigating} on the 19LA \cite{asvspoof19} dataset. The proposed ASR pretrained Conformer model, despite having only 13M parameters, outperforms two Wav2Vec 2.0 front-end models that have over 300M parameters. Moreover, when compared to the best-performing single-system smaller model, AASIST, the Conformer model exhibits faster training and inference speeds due to its lack of a complex graph neural network structure.
\begin{table}[t]
\caption{Comparison with recently proposed state-of-the-art systems, reported using pooled EER (\%) on 19LA evaluation set. Systems are displayed in an ascending order using the number model parameters. The $\dagger$ model is implemented without any data augmentation. }
\label{tab:19LAeer}
\resizebox{\linewidth}{!}{%
\begin{tabular}{llll}
\hline
System     & \# Param  & Architecture             & EER \\ \hline
Jung et al. \cite{aasist} & 297K   & AASIST                   & 0.83 \\
Tak et al. \cite{tak21_rawgat}       & 437K   & RawGAT-ST                & 1.06 \\
Zhang et al. \cite{zhang21da_interspeech}    & 1,100K & SENet                    & 1.14 \\
\textbf{Ours}       & 13M  & ASR pretrained Conformer & \textbf{0.72} \\
\textbf{Ours}       & 13M  & ASV pretrained Conformer & 1.31 \\
Lee et al. \cite{lee22q_interspeech}       & 300+M  & W2V(XLSR-53)+ASP         & \textbf{0.31} \\
Tak et al. \cite{tak2022automatic} $\dagger$        & 300+M  & W2V+AASIST      & \textbf{0.37} \\
Wang et al. \cite{wang2021investigating}     & 300+M  & W2V(Large2)+LLGF         & 0.86 \\
Wang et al.  \cite{wang2021investigating}     & 300+M  & W2V(XLSR-53)+LGF         & 1.28 \\ \hline
\end{tabular}
}
\end{table}

\subsection{Comparison with other Conformer-based CM systems}
Table~\ref{tab:19LAeer_Conformer} presents a comparison of the proposed Conformer model and the performance of other Conformer-based CM systems on 19LA dataset. Rosello et al. \cite{rosello22_iberspeech} developed CM systems by utilizing classification tokens as output features and linking FC layers or decoders in the backend. However, as indicated in Table~\ref{tab:19LAeer_Conformer}, training of the Conformer model directly on small-scale anti-spoofing data is vulnerable to overfitting, resulting in degradation of the generalization performance. Pre-training, on the other hand, can expedite model fitting and enhance model robustness.
\begin{table}[t]
\caption{Comparison with other Conformer-based CM systems, reported using pooled EER (\%) on 19LA evaluation set.}
\center
\label{tab:19LAeer_Conformer}
\resizebox{0.9\linewidth}{!}{%
\begin{tabular}{lll}
\hline
System  & Architecture             & EER \\ \hline
Rosello \cite{rosello22_iberspeech} & Conformer + Decoder2     & 7.51 \\
\multirow{3}{*}{\textbf{Ours}}    & W/O pretraining          & 6.06 \\
    & ASR pretrained Conformer & \textbf{0.72} \\
   & ASV pretrained Conformer & 1.31 \\ \hline
\end{tabular}}
\end{table}
\section{Conclusion}

In this paper, we proposed a CM system based on transfer learning with ASR or ASV pre-trained MAF-Conformer constructs. We validated the effectiveness of the proposed method on two different language's anti-spoofing databases, FAD and ASVspoof. Our results demonstrate that the pretrained model converges faster and performs better compared to directly training a Conformer model on anti-spoofing database. Furthermore, when compared to LFCC and ResNet34 models, the ASR pre-trained Conformer model consistently achieves significantly better results on each database. In addition, we evaluated the robustness of different models against various spoofing algorithms on the ASVspoof 2019 LA evaluation set. Our findings clarify that the performance of neural network-based CM systems is not solely correlated with whether or not they have seen a spoofing algorithm in training. We propose ET metrics for measuring the robustness of models to certrain spoofing algorithms. These metrics may be useful for model fusion and feature selection for spoofing algorithm traceability tasks.
In our future work, we will explore the fusion of ASV and ASR pre-trained Conformer models along three dimensions: embedding, score, and model parameters.

\bibliographystyle{IEEEbib}
\bibliography{template}

\end{document}